# A Molecular Model for Communication through a Secrecy System.

Okunoye Babatunde O.
Department of Pure and Applied Biology, Ladoke Akintola University of Technology,
P.M.B 4000 Ogbomoso, Nigeria. **Email:** babatundeokunoye@yahoo.co.uk

Codes have been used for centuries to convey secret information. To a cryptanalyst, the interception of a code is only the first step in recovering a secret message. Deoxyribonucleic acid (DNA) is a biological and molecular code. Through the work of Marshall Nirenberg and others, DNA is now understood to specify for amino acids in triplet codes of bases. The possibility of DNA encoding secret information in a natural language is explored, since a code is expected to have a distinct mathematical solution.

## INTRODUCTION

The problems of cryptography and secrecy systems furnish an interesting application of communication theory (Shannon, 1948). A secrecy system can be defined as a single transformation on a language of the form $T_1 M = E$ where $T_1$ = transformation, M = Message and E = Cryptogram (Shannon, 1949). A code carrying secret information is expected to have a distinct mathematical solution. Codes have extensive military applications, and the breaking of the Enigma code was pivotal in the Allied War effort in the Second World War.

Marshall Nirenberg (Nirenberg & Matthei, 1961; Nirenberg & Leder, 1964) and co-workers at the National Institutes of Health Bethesda, Maryland, U.S.A. between the years 1961-1964 put forward a biochemical solution to the genetic code. We now understand that four bases of DNA are arranged as 64 triplet codons, which in turn specify for 20 amino acids (Brock & Madigan, 1991). A mathematical solution to the genetic code is proposed in this paper, using the virus Bacteriophage T4 as a model.

## METHODS

The numbers of Adenine, Thymine, Guanine and Cytosine residues were counted in 3,183 turns of bases in the genome of Bacteriophage T4, in the 5' to 3' direction (i.e from base 168900 backwards). This represents 31,830 bases. Bacteriophge T4 and its genome sequence represents the best understood model for functional genomics and proteomics (Miller et al., 2003). Bacteriophage T4 base sequence was obtained for GenBank with accession number AF158101.

There are 10 bases per turn of the DNA helix. (Brock & Madigan, 1991). There fore the numbers of Adenine, Thymine, Guanine and Cytosine which Per turn of the helix would add up to 10 will have different permutations. For Example Adenine 1, Thymine 2, Guanine 3, Cytosine 4. Another example is Adenine 0, Thymine 5, Guanine 0, Cytosine 5. Throughout the course of investigation involving 3,183 turns of bases, no individual base occurred more than 8 times. The number of any one bases ranged from 0,1,2,3,… 8. A base which did not appear per turn was recorded as 0. Thus we have 230 permutations of the numbers 0,1,2,3,….8 which add up to 10. These 230 permutations can be grouped into 21 groups based on the numbers which constitute them. For example, the permutations 0055,0505, 5005, 0550, 5500, 5050 are grouped as one group. The frequencies are probabilities of each number group in 3,183 turns of T4 phage DNA was calculated.

Assuming that the genetic code is producing English text, the probabilities of English letters were calculated in 3,183 letters chosen from chapter 5 of Wuthering Heights (Brönte, 1965), a classical English novel. Considerable reductions in text are possible in the English Language without losing information due to the statistical nature of the language, high frequency of certain words etc. (Shannon, 1949). This property of the English Language whereby certain letters can be omitted without losing information conveyed in the language is called Redundancy (Shannon, 1949). Redundancy is of central importance in the study of secrecy systems (Shannon, 1949). The letters C, Q, V, X, Z, were thus omitted. The frequency histogram of both the number groups and English letters show similarities. A simple substitution is then established, by replacing each number group with English letters, in order of increasing probabilities.

**RESULTS**

When the substitution is effected, 3183 English letters are obtained. The probable word method (Shannon, 1949) is used to recover any secret message in the English text produced by Substitution. The 'probable words' may be words or phrases expected in the particular message due to its source, or they may merely be

| S/N | Number Group | Number of Permutations | Frequency | Probability |
|---|---|---|---|---|
| 1 | 0055(X18) | 6 | 5 | 0.0016 |
| 2 | 0028(X19) | 12 | 7 | 0.0022 |
| 3 | 0118(X21) | 12 | 11 | 0.0035 |
| 4 | 1117(X14) | 4 | 15 | 0.0047 |
| 5 | 0037(X17) | 12 | 16 | 0.0050 |
| 6 | 0046(X20) | 12 | 23 | 0.0072 |
| 7 | 0226(X16) | 12 | 63 | 0.0198 |
| 8 | 0127(X3) | 16 | 76 | 0.0239 |
| 9 | 0136(X10) | 16 | 109 | 0.0343 |
| 10 | 1144(X13) | 6 | 119 | 0.0374 |
| 11 | 0244(X6) | 12 | 135 | 0.0424 |
| 12 | 0145(X1) | 16 | 140 | 0.0440 |
| 13 | 1126(X8) | 12 | 144 | 0.0453 |
| 14 | 0334(X12) | 12 | 145 | 0.0456 |
| 15 | 1333(X9) | 4 | 181 | 0.0569 |
| 16 | 2224(X15) | 4 | 188 | 0.0591 |
| 17 | 1135(X7) | 12 | 195 | 0.0613 |
| 18 | 0235(X11) | 16 | 225 | 0.0707 |
| 19 | 1225(X4) | 12 | 290 | 0.0911 |
| 20 | 2233(X2) | 6 | 295 | 0.0927 |
| 21 | 1234(X5) | 16 | 800 | 0.2514 |
|  |  | 230 | 3183 | 0.9988 |

**Figure 1a.** Table showing the number groups, the number of their permutations, their frequencies and probabilities in 3,183 turns of T4 Phage Genome.

| S/N | LETTER | FREQUENCY | PROBABILITY |
|---|---|---|---|
| 1 | J | 5 | 0.0016 |
| 2 | K | 26 | 0.0082 |
| 3 | P | 52 | 0.0163 |
| 4 | F | 53 | 0.0167 |
| 5 | G | 69 | 0.0217 |
| 6 | Y | 70 | 0.0220 |
| 7 | B | 83 | 0.0261 |
| 8 | W | 86 | 0.0270 |
| 9 | M | 91 | 0.0286 |
| 10 | U | 92 | 0.0289 |
| 11 | L | 131 | 0.0412 |
| 12 | D | 149 | 0.0468 |
| 13 | R | 183 | 0.0575 |
| 14 | O | 219 | 0.0688 |
| 15 | S | 221 | 0.0694 |
| 16 | N | 235 | 0.0738 |
| 17 | T | 239 | 0.0751 |
| 18 | H | 253 | 0.0795 |
| 19 | I | 253 | 0.0795 |
| 20 | A | 291 | 0.0914 |
| 21 | E | 382 | 0.1200 |
|  |  | 3183 | 1.0001 |

**Figure 1b.** Frequencies and Probabilities of English Letter out of 3,183 letters of an English text (Wuthering Heights) in the case of the letters H and I which both occur 253 times, thereby having the same probabilities, the consonant is placed before the vowel.

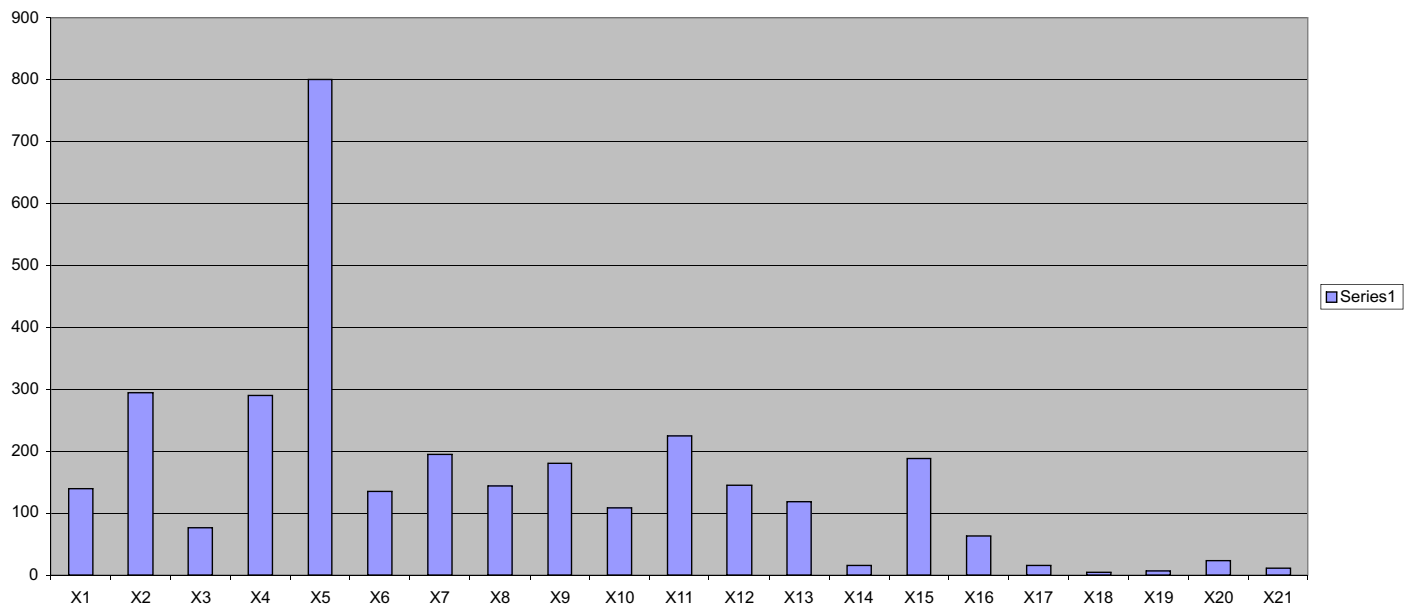

**Frequency graph of number groups in 3,183 turns of T4 phage DNA. The number groups are written in the order they appear in T4 phage DNA in the 5' to 3' direction.**

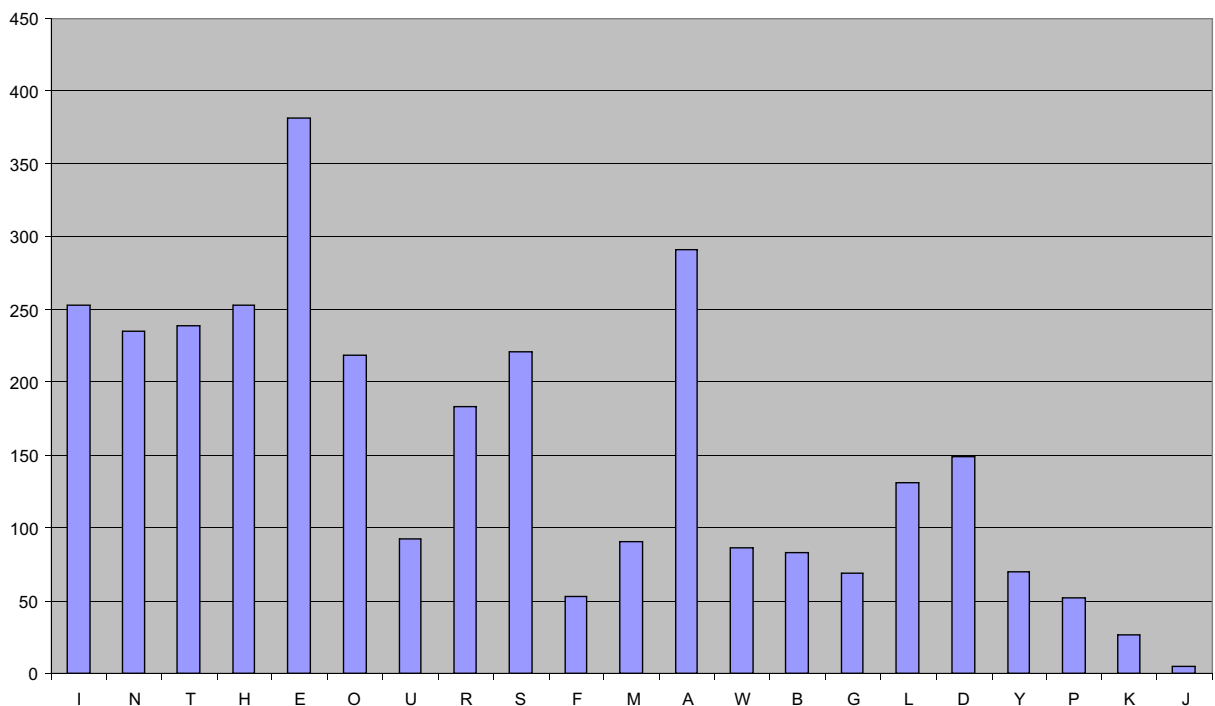

**Frequency graph of 3,183 letters of chapter 5 of Wuthering Heights. The letters are arranged in the order they appear in the text.**

| S/N | Number group | English Letter |
|-----|--------------|----------------|
| 1   | 0055         | J              |
| 2   | 0028         | K              |
| 3   | 0118         | P              |
| 4   | 1117         | F              |
| 5   | 0037         | G              |
| 6   | 0046         | Y              |
| 7   | 0226         | B              |
| 8   | 0127         | W              |
| 9   | 0136         | M              |
| 10  | 1144         | U              |
| 11  | 0244         | L              |
| 12  | 0145         | D              |
| 13  | 1126         | R              |
| 14  | 0334         | O              |
| 15  | 1333         | S              |
| 16  | 2224         | N              |
| 17  | 1135         | T              |
| 18  | 0235         | H              |
| 19  | 1225         | I              |
| 20  | 2233         | A              |
| 21  | 1234         | E              |

Figure 2. A simple substitution table

Words or phrases expected in the particular message due to its source, or they may merely be common words or syllables which occur in any text in the language, such as *the, and, tion, that,* and the like in English (Shannon, 1949).

Using the probable word method did not exactly yield a message, yet it was discovered that over 300 English words could be spelt out from the text. Phrases occur occasionally, and it was Possible to make out a few sentences from words adjacent to each other or closely aggregated. Reconstructions are necessary to bring out the sense in the sentences. The only sentence found not needing reconstruction was *HO A SEAL* which, is an exclamation *HO A SEAL!*

Hundreds of words and phrases were found in the text without reconstruction; nevertheless if reconstructions are applied to words and phrases, the word and phrase count will be considerably increased.

| S/N | WORDS | PHRASES | SENTENCES |
|---|---|---|---|
| 1 | IT | 'BE HIM' | HO A SEAL. |
| 2 | HE | 'A SEA' | **I** EE **TIRE** = I TIRE |
| 3 | IS | 'IT DIE' | **WE** RRH **HE EAT** EE **A TREE** = WE EAT A TREE/HE EAT A TREE. |
| 4 | ME | 'LET TWO' | **USE** O**HE A** NE **SEA** = HE USE A SEA |
| 5 | US | 'I READ' | **HE A IS** RR **WOE** = HE IS A WOE/ HE IS WOE. |
| 6 | THEIR | 'AN HAT' | **WOE** E **HIT** YA **IT** = WOE HIT IT. |
| 7 | DEN | 'AS IS' | **A** W **NUN HE** E **SEE** = A NUN HE SEE(S). |
| 8 | SUN | 'SEE A SANE' | **TOE LET** AU **TIMS** RTRI **HEAL** = LET TIM'S TOE HEAL. |
| 9 | MEN | 'A RIDE' | **HEAL** DI **US** = HEAL US. |
| 10 | ROB | 'LORE DEN' | **SO** E **LOAD** ET **LET TWO** = SO LET TWO LOAD. |
| 11 | LIE | 'USE LEG' | **I READ I**HT **SIR** = I READ (IT) SIR. |
| 12 | SEED | 'END IN' | **IT** T **FIT HUE** = HUE IT FIT. |
| 13 | AREA | 'HER HAT' | **RED EROS** H **BE** B **ME** = RED EROS BE ME / RED ROSE BE ME. |
| 14 | HEAL | 'SELL BEAR' | **YES** HH **HE ERED** = YES HE ERRED. |
| 15 | WEST | 'BE OR' | **HE** A **SEE ASA NET** = HE SEE(S) A NET/ HE SEES AS A NET. |
| 16 | STAR | 'I SUE' | |
| 17 | ELITE | 'NUNS ARE' | |
| 18 | HEARD | 'A REAL' | |
| 19 | TENET | 'LET TWO' | |
| 20 | BERTH | 'TEA SEED' | |

**Figure 3a.** Some words, phrases and sentences found in the text.

| S/N | WORDS AND PHRASES | RECONSTRUCTION |
|---|---|---|
| 1 | TWROH | THROW |
| 2 | REAML | REALM |
| 3 | STARDMO | STARDOM |
| 4 | SEDN | SEND |
| 5 | SEALDE | SEALED |
| 6 | EROS[†] | ROSE |
| 7 | EAEST | EAST |
| 8 | ADEEENOSINE | ADENOSINE |
| 9 | HEAEEARD | HEARD |
| 10 | SEEED | SEED |
| 11 | NEEAT | NEAT |
| 12 | HEEAT | HEAT/HE EAT |
| 13 | TEEEN | TEN/TEEN |
| 14 | HEAET | HEAT/HE ATE |
| 15 | ROTEN | ROTTEN |
| 16 | ISAUED AT | ISSUED AT |
| 17 | INHELAED | IN HE LAID |
| 18 | SUNGER SOLD | SINGER SOLD |
| 19 | SADEN | SADDEN |
| 20 | BEBLE | BIBLE |

Figure 3b: Some additional words and phrases by reconstruction.

[†]. EROS is a Latin word in its own right, meaning love

**DISCUSSION**

In substituting the number groups with English Letters, the assumption made was that the genetic code is producing English text. For the 3,183 turns of T4 phage DNA investigated, no individual base occurred more than 8 times. This produces 21 number groups. Had any of the bases exceeded 8, e.g. 9, an additional number group consisting of the permutations of the numbers 0, 0, 1, 9 will be added, making 22. Had any of the bases exceeded 9, e.g. 10, the number groups will total 23, because of the permutations of the numbers 0,0,0,10.

It remains to be shown whether a similar result can be obtained with other languages, especially the ones with number of letters nearing 21, for example Hebrew with 22 letters/alphabets.

Counting of number groups and substitution with English letters was done in the 5' to 3' direction, i.e. starting from the last turn of T4 phage DNA (168900 168891) backwards. A similar result might be obtained in the 3' to 5' direction.

Finally, in choosing which of the English letters will be omitted, the author did not rely solely on the property Redundancy (D). For example the letter *u* usually follows *q* in English words, so the *u* can be omitted without loss (Shannon, 1949). Yet the letter *u* was not omitted. It can be said that intuition played a role in choosing what to include and omit. It also remains to be seen if a similar result can be obtained by omitting a different set of letters of the English alphabet.